\tolerance=500
%
\catcode`@=11 
%
%
%

\font\fourteenrm=cmr10 scaled\magstep2
\font\twelverm=cmr10 scaled\magstep1
\font\ninerm=cmr9
\font\eightrm=cmr8
\font\sixrm=cmr6

\font\fourteenbf=cmbx10 scaled\magstep2
\font\twelvebf=cmbx10 scaled\magstep1
\font\ninebf=cmbx9
\font\eightbf=cmbx8
\font\sixbf=cmbx6
\font\seventeeni=cmmi10 scaled\magstep3     \skewchar\seventeeni='177
\font\fourteeni=cmmi10 scaled\magstep2      \skewchar\fourteeni='177
\font\twelvei=cmmi10 scaled\magstep1        \skewchar\twelvei='177
\font\ninei=cmmi9                           \skewchar\ninei='177
\font\eighti=cmmi8                          \skewchar\eighti='177
\font\sixi=cmmi6                            \skewchar\sixi='177
\font\seventeensy=cmsy10 scaled\magstep3    \skewchar\seventeensy='60
\font\fourteensy=cmsy10 scaled\magstep2     \skewchar\fourteensy='60
\font\twelvesy=cmsy10 scaled\magstep1       \skewchar\twelvesy='60
\font\ninesy=cmsy9                          \skewchar\ninesy='60
\font\eightsy=cmsy8                         \skewchar\eightsy='60
\font\sixsy=cmsy6                           \skewchar\sixsy='60

\font\fourteenex=cmex10 scaled\magstep2
\font\twelveex=cmex10 scaled\magstep1

\font\fourteensl=cmsl10 scaled\magstep2
\font\twelvesl=cmsl10 scaled\magstep1
\font\ninesl=cmsl9
\font\eightsl=cmsl8

\font\fourteenit=cmti10 scaled\magstep2
\font\twelveit=cmti10 scaled\magstep1
\font\nineit=cmti9
\font\eightit=cmti8
\font\twelvett=cmtt10 scaled\magstep1
\font\ninett=cmtt9
\font\eighttt=cmtt8
\font\twelvecp=cmcsc10 scaled\magstep1
\font\tencp=cmcsc10
\newfam\cpfam
%
%
%
%
\fontdimen16\seventeensy=4.59pt \fontdimen17\seventeensy=4.59pt
\fontdimen16\fourteensy=3.78pt \fontdimen17\fourteensy=3.78pt
\fontdimen16\twelvesy=3.24pt \fontdimen17\twelvesy=3.24pt
\fontdimen16\tensy=2.70pt \fontdimen17\tensy=2.70pt
\fontdimen16\ninesy=2.43pt \fontdimen17\ninesy=2.43pt
\fontdimen16\eightsy=2.16pt \fontdimen17\eightsy=2.16pt
\fontdimen16\sixsy=1.62pt \fontdimen17\sixsy=1.62pt
%
%
%
\newcount\f@ntkey \f@ntkey=0
\def\samef@nt{\relax\ifcase \f@ntkey \rm \or\oldstyle \or\or
  \or\it \or\sl \or\bf \or\tt \or\caps \fi}
%
%
%
\def\eightpoint{\relax
  \textfont0=\eightrm \scriptfont0=\sixrm \scriptscriptfont0=\fiverm
  \textfont1=\eighti \scriptfont1=\sixi \scriptscriptfont1=\fivei
  \textfont2=\eightsy \scriptfont2=\sixsy \scriptscriptfont2=\fivesy
  \textfont3=\tenex \scriptfont3=\tenex \scriptscriptfont3=\tenex
  \textfont\itfam=\eightit
  \textfont\slfam=\eightsl
  \textfont\bffam=\eightbf \scriptfont\bffam=\sixbf
    \scriptscriptfont\bffam=\fivebf
  \textfont\ttfam=\eighttt
  \textfont\cpfam=\tencp
  \def\rm{\fam0 \eightrm \f@ntkey=0 }
  \def\oldstyle{\fam1 \eighti \f@ntkey=1 }
  \def\it{\fam\itfam \eightit \f@ntkey=4 }
  \def\sl{\fam\slfam \eightsl \f@ntkey=5 }
  \def\bf{\fam\bffam \eightbf \f@ntkey=6 }
  \def\tt{\fam\ttfam \eighttt \f@ntkey=7 }
  \def\caps{\fam\cpfam \tencp \f@ntkey=8 }
  \h@big=6.8\p@{} \h@Big=9.2\p@{} \h@bigg=11.6\p@{} \h@Bigg=14\p@{}
  \setbox\strutbox=\hbox{\vrule height 6.8pt depth 2.8pt width\z@}
  \samef@nt}
\def\ninepoint{\relax
  \textfont0=\ninerm \scriptfont0=\sixrm \scriptscriptfont0=\fiverm
  \textfont1=\ninei \scriptfont1=\sixi \scriptscriptfont1=\fivei
  \textfont2=\ninesy \scriptfont2=\sixsy \scriptscriptfont2=\fivesy
  \textfont3=\tenex \scriptfont3=\tenex \scriptscriptfont3=\tenex
  \textfont\itfam=\nineit
  \textfont\slfam=\ninesl
  \textfont\bffam=\ninebf \scriptfont\bffam=\sixbf
    \scriptscriptfont\bffam=\fivebf
  \textfont\ttfam=\ninett
  \textfont\cpfam=\tencp
  \def\rm{\fam0 \ninerm \f@ntkey=0 }
  \def\oldstyle{\fam1 \ninei \f@ntkey=1 }
  \def\it{\fam\itfam \nineit \f@ntkey=4 }
  \def\sl{\fam\slfam \ninesl \f@ntkey=5 }
  \def\bf{\fam\bffam \ninebf \f@ntkey=6 }
  \def\tt{\fam\ttfam \ninett \f@ntkey=7 }
  \def\caps{\fam\cpfam \tencp \f@ntkey=8 }
  \h@big=7.65\p@{} \h@Big=10.35\p@{} \h@bigg=13.05\p@{} \h@Bigg=15.75\p@{}
  \setbox\strutbox=\hbox{\vrule height 7.65pt depth 3.15pt width\z@}
  \samef@nt}
\def\tenpoint{\relax
  \textfont0=\tenrm \scriptfont0=\sevenrm \scriptscriptfont0=\fiverm
  \textfont1=\teni \scriptfont1=\seveni \scriptscriptfont1=\fivei
  \textfont2=\tensy \scriptfont2=\sevensy \scriptscriptfont2=\fivesy
  \textfont3=\tenex \scriptfont3=\tenex \scriptscriptfont3=\tenex
  \textfont\itfam=\tenit
  \textfont\slfam=\tensl
  \textfont\bffam=\tenbf \scriptfont\bffam=\sevenbf
    \scriptscriptfont\bffam=\fivebf
  \textfont\ttfam=\tentt
  \textfont\cpfam=\tencp
  \def\rm{\fam0 \tenrm \f@ntkey=0 }
  \def\oldstyle{\fam1 \teni \f@ntkey=1 }
  \def\it{\fam\itfam \tenit \f@ntkey=4 }
  \def\sl{\fam\slfam \tensl \f@ntkey=5 }
  \def\bf{\fam\bffam \tenbf \f@ntkey=6 }
  \def\tt{\fam\ttfam \tentt \f@ntkey=7 }
  \def\caps{\fam\cpfam \tencp \f@ntkey=8 }
  \h@big=8.5\p@{} \h@Big=11.5\p@{} \h@bigg=14.5\p@{} \h@Bigg=17.5\p@{}
  \setbox\strutbox=\hbox{\vrule height 8.5pt depth 3.5pt width\z@}
  \samef@nt}
\def\twelvepoint{\relax
  \textfont0=\twelverm \scriptfont0=\ninerm \scriptscriptfont0=\sixrm
  \textfont1=\twelvei \scriptfont1=\ninei \scriptscriptfont1=\sixi
  \textfont2=\twelvesy \scriptfont2=\ninesy \scriptscriptfont2=\sixsy
  \textfont3=\twelveex \scriptfont3=\twelveex \scriptscriptfont3=\twelveex
  \textfont\itfam=\twelveit
  \textfont\slfam=\twelvesl \scriptfont\slfam=\ninesl
  \textfont\bffam=\twelvebf \scriptfont\bffam=\ninebf
    \scriptscriptfont\bffam=\sixbf
  \textfont\ttfam=\twelvett
  \textfont\cpfam=\twelvecp
  \def\rm{\fam0 \twelverm \f@ntkey=0 }
  \def\oldstyle{\fam1 \twelvei \f@ntkey=1 }
  \def\it{\fam\itfam \twelveit \f@ntkey=4 }
  \def\sl{\fam\slfam \twelvesl \f@ntkey=5 }
  \def\bf{\fam\bffam \twelvebf \f@ntkey=6 }
  \def\tt{\fam\ttfam \twelvett \f@ntkey=7 }
  \def\caps{\fam\cpfam \twelvecp \f@ntkey=8 }
  \h@big=10.2\p@{} \h@Big=13.8\p@{} \h@bigg=17.4\p@{} \h@Bigg=21.0\p@{}
  \setbox\strutbox=\hbox{\vrule height 10pt depth 4pt width\z@}
  \samef@nt}
\def\fourteenpoint{\relax
  \textfont0=\fourteenrm \scriptfont0=\tenrm \scriptscriptfont0=\sevenrm
  \textfont1=\fourteeni \scriptfont1=\teni \scriptscriptfont1=\seveni
  \textfont2=\fourteensy \scriptfont2=\tensy \scriptscriptfont2=\sevensy
  \textfont3=\fourteenex \scriptfont3=\fourteenex
    \scriptscriptfont3=\fourteenex
  \textfont\itfam=\fourteenit
  \textfont\slfam=\fourteensl \scriptfont\slfam=\tensl
  \textfont\bffam=\fourteenbf \scriptfont\bffam=\tenbf
    \scriptscriptfont\bffam=\sevenbf
  \textfont\ttfam=\twelvett
  \textfont\cpfam=\twelvecp
  \def\rm{\fam0 \fourteenrm \f@ntkey=0 }
  \def\oldstyle{\fam1 \fourteeni \f@ntkey=1 }
  \def\it{\fam\itfam \fourteenit \f@ntkey=4 }
  \def\sl{\fam\slfam \fourteensl \f@ntkey=5 }
  \def\bf{\fam\bffam \fourteenbf \f@ntkey=6 }
  \def\tt{\fam\ttfam \twelvett \f@ntkey=7 }
  \def\caps{\fam\cpfam \twelvecp \f@ntkey=8 }
  \h@big=11.9\p@{} \h@Big=16.1\p@{} \h@bigg=20.3\p@{} \h@Bigg=24.5\p@{}
  \setbox\strutbox=\hbox{\vrule height 12pt depth 5pt width\z@}
  \samef@nt}
%
%
%
\newdimen\h@big
\newdimen\h@Big
\newdimen\h@bigg
\newdimen\h@Bigg
\def\big#1{{\hbox{$\left#1\vbox to\h@big{}\right.\n@space$}}}
\def\Big#1{{\hbox{$\left#1\vbox to\h@Big{}\right.\n@space$}}}
\def\bigg#1{{\hbox{$\left#1\vbox to\h@bigg{}\right.\n@space$}}}
\def\Bigg#1{{\hbox{$\left#1\vbox to\h@Bigg{}\right.\n@space$}}}
%
%
%
\newskip\normaldisplayskip
\newskip\normaldispshortskip
\newskip\normalparskip
\newskip\normalskipamount
\normalbaselineskip = 20pt plus 0.2pt minus 0.1pt
\normallineskip = 1.5pt plus 0.1pt minus 0.1pt
\normallineskiplimit = 1.5pt
\normaldisplayskip = 20pt plus 5pt minus 10pt
\normaldispshortskip = 6pt plus 5pt
\normalparskip = 6pt plus 2pt minus 1pt
\normalskipamount = 5pt plus 2pt minus 1.5pt
%
%
\def\sp@cing#1{%
  \baselineskip=\normalbaselineskip%
    \multiply\baselineskip by #1%
    \divide\baselineskip by 12%
  \lineskip=\normallineskip%
    \multiply\lineskip by #1%
    \divide\lineskip by 12%
  \lineskiplimit=\normallineskiplimit%
    \multiply\lineskiplimit by #1%
    \divide\lineskiplimit by 12%
  \parskip=\normalparskip%
    \multiply\parskip by #1%
    \divide\parskip by 12%
  \abovedisplayskip=\normaldisplayskip%
    \multiply\abovedisplayskip by #1%
    \divide\abovedisplayskip by 12%
  \belowdisplayskip=\abovedisplayskip%
  \abovedisplayshortskip=\normaldispshortskip%
    \multiply\abovedisplayshortskip by #1%
    \divide\abovedisplayshortskip by 12%
  \belowdisplayshortskip=\abovedisplayshortskip%
    \advance\belowdisplayshortskip by \belowdisplayskip%
    \divide\belowdisplayshortskip by 2%
  \smallskipamount=\normalskipamount%
    \multiply\smallskipamount by #1%
    \divide\smallskipamount by 12%
  \medskipamount=\smallskipamount%
    \multiply\medskipamount by 2%
  \bigskipamount=\smallskipamount%
    \multiply\bigskipamount by 4}
%
%
%
\newcount\fontsize
\def\Eightpoint{\eightpoint\fontsize=8\sp@cing{8}}
\def\Ninepoint{\ninepoint\fontsize=9\sp@cing{9}}
\def\Tenpoint{\tenpoint\fontsize=10\sp@cing{10}}
\def\Twelvepoint{\twelvepoint\fontsize=12\sp@cing{12}}
\def\Fourteenpoint{\fourteenpoint\fontsize=14\sp@cing{14}}
\newcount\spacesize
\def\singlespace{\spacesize=\fontsize
  \multiply\spacesize by 9\divide\spacesize by 12%
  \sp@cing{\spacesize}}
\def\normalspace{\sp@cing{\fontsize}}
\def\doublespace{\spacesize=\fontsize
  \multiply\spacesize by 15\divide\spacesize by 12%
  \sp@cing{\spacesize}}
\Twelvepoint 
\interlinepenalty=50
\interfootnotelinepenalty=5000
\predisplaypenalty=9000
\postdisplaypenalty=500
\hfuzz=1pt
\vfuzz=0.2pt
\catcode`@=12 
\hsize=16 cm
\vsize=24 cm
\parskip=0mm  
\line{}
\vskip 2mm

\Fourteenpoint 

{\centerline{\bf The production model of Ishida {\it et al.}}}
{\centerline{\bf and unitarity}}
\vskip 1cm

{\centerline{M.R. Pennington}}
\vskip 1mm

\Twelvepoint  
\centerline{Centre for Particle Theory,
University of Durham,}

\centerline{Durham DH1 3LE, U.K.}
\vskip 1cm
\Twelvepoint
\centerline{ABSTRACT}
\vskip 5mm
\baselineskip=5mm   
\parskip=0mm  
{\leftskip 1.5cm\rightskip 1.5cm{\noindent The relation between scattering and
production amplitudes imposed by unitarity and analyticity, recently
criticised by Ishida {\it et al.}~$^{1),2)}$, is explained.}\par}
\vskip 1cm

\Twelvepoint
\baselineskip=7mm
\parskip=2mm
In a contribution to the Seventh International Conference on Hadron Spectroscopy
at Brookhaven,
Ishida { \it et al.}~$^{1),2)}$ have questioned one of the orthodox methods of 
implementing the final state interaction theorem
in production processes.  Here we show that this criticism is incorrect, being based on a misunderstanding of the method.
The case considered by Ishida {\it et al.}~$^{1),2),3)}$ is particularly simple, it is that of
a single channel,
for instance $\pi\pi\to\pi\pi$. It is then well-known that each unitary
 partial wave amplitude  can, for real $s$ ---
the square of the c.m. energy, be represented by
$${\cal T}(s)\,=\, {K(s)\over{1-i\rho(s) K(s)}}\quad ,\eqno(1)$$
where $\rho(s)$ is the standard phase-space factor of $\, 2p/\sqrt{s}\,$ ($p$ being the
c.m. 3-momentum) and $K(s)$ is real.  Importantly,
$K(s)$ embodies any real zeros of the amplitude ${\cal T}(s)$.

Now Watson's final state interaction theorem~$^{4)}$ requires that any other (non-strongly interacting)
process producing the same final state must have its corresponding partial wave
${\cal F}(s)$ having the same phase.
To implement this within the $K$--matrix formalism, Aitchison~$^{5)}$ proposed representing
${\cal F}(s)$ by
$${\cal F}(s)\,=\,{P(s)\over{1-i\rho(s) K(s)}}\quad ,\eqno(2)$$
where the function $P(s)$, like $K(s)$, is real for real values of $s$. 
The complex denominator, $1-i\rho K$, not only ensures the production
amplitude has the same phase as the elastic one, Eq.~(1) for $s$ real, but also
ensures  that physical states, which are poles in the complex $s$--plane on 
the nearby unphysical sheets, transmit from one process to another
 through this universal denominator.
 
  It is common practice to parameterise the $K$--matrix in
terms of poles.
However, these introduce artificial zeros, Eq.~(2), in the production amplitude, unless the function $P(s)$ has the very same poles.  A simple method of implementing
this constraint has been proposed by AMP~$^{6)}$.  This is to express the $P$--vector as
$$P(s)\,=\,\alpha(s)\, {\hat K}(s)\quad ,\eqno(3)$$ 
where ${\hat K}$ is the reduced $K$--matrix, which contains the poles of $K(s)$,
but with its zeros divided out.  In this simple
representation
$${\cal F}(s)\,=\,\alpha(s)\, {\hat{\cal T}}(s) \eqno(4)$$
with ${\hat{\cal T}}(s)$ being the $T$-matrix with its zeros removed~$^{6)}$.
In general, then analyticity requires $\alpha(s)$ to be a {\it smooth} function for
$s > s_{threshold}$~$^{6),7)}$.  In physical $\pi\pi$ scattering, the only such zeros
to divide out are the Adler zeros below threshold for the $S$--waves
 and the kinematic zeros at threshold for higher partial waves.
 Such zeros are divided out, since zeros of the amplitudes
${\cal T}(s)$ will not, in general, transmit to production amplitudes, ${\cal F}(s)$, though of course poles do.

Ishida {\it et al.}~$^{2)}$ construct a simple one-channel model in which
the $K$--matrix has two poles, so that
$$K(s)\,=\, {g_1^2\over{s-m_1^2}}\,+\,{g_2^2\over{s-m_2^2}}\; .\eqno(5)$$
Clearly, this example has a zero at $s=s_0$ between $s=m_1^2$ and $s=m_2^2$, where
$$s_0\,=\,\left(g_1^2 m_2^2 + g_2^2 m_1^2\right)/(g_1^2 + g_2^2)\quad .\eqno(6)$$
This zero in ${\cal T}$ will not, in general, occur in 
production processes.  Hence, it is necessary to define a reduced 
${\hat{\cal T}}$ or equivalently
${\hat K}$--matrix.  Then any production process can be expressed as
$${\cal F}(s)\,=\,\alpha(s)\, {{\hat K(s)}\over {1-i\rho(s) K(s)}}\quad ,\eqno(7) $$
where
$${\hat K}(s)\,=\,{(g_1^2+g_2^2)\over{(s-m_1^2)(s-m_2^2)}}\eqno(8)$$
and $\alpha(s)$ having only a left hand cut is expected to be a smooth function
for $s > s_{threshold}$.  If the reduced ${\hat K}$--matrix is not used, but instead $K$ itself, as in the example of
Ishida {\it et al.}~$^{2)}$, a spurious zero transmits to the production
process, unphysically shackling its description.  

This discussion is readily generalised to $n$ coupled channels, when ${\cal T}$
and $K$ are $n\times n$ matrices and ${\cal F}$ and $\alpha$ are 
$n$--component vectors. However, when there is only one channel, the reason for introducing the
reduced ${\hat K}$--matrix is particularly transparent. The hadronic amplitude ${\cal T}$ can be written
in terms of the phase-shift $\delta$ as
$${\cal T}(s)\,=\,{1\over{\rho}}\, \sin \delta\,e^{i\delta}\quad. \eqno(9)$$
The $K$--matrix element is then $\tan\delta/\rho$.  Clearly, the
$K$--matrix has poles when $\delta = (2n+1)\pi/2$ (with $n=1,2,...$)
and the amplitude ${\cal T}$ has zeros when $\delta = n\pi$ (again with $n$ an integer).
In terms of the phase-shift, the production amplitude
${\cal F}$ of Eq.~(2) becomes 
$${\cal F}(s)\;=\;{1\over \rho}\, P \cos \delta \,e^{i\delta}\quad . \eqno(10)$$
It is then obvious that unless $P(s)$ has the poles of the $K$--matrix,
${\cal F}(s)$ will be zero exactly where resonances are expected to show up,
 {\it i.e.}  when $\delta = (2n+1)\pi/2$ making
$\cos \delta =0$. Choosing
$P(s)$ to be simply proportional to $K(s)$ replaces $\cos \delta$ in Eq.~(10) by 
$\sin \delta$. However, then  the zeros of ${\cal T}$,
Eq.~(9),
at $\delta = n\pi$, unnecessarily  transmit to the
production amplitude.  Thus these zeros must be removed by defining
the reduced $K$--matrix, ${\hat K}$, as
$${\hat K}(s)\;=\;K(s)/\prod_n (s-s_n)\; ,\eqno(11)$$
where $\delta(s_n)\,=\,n\pi$,  so that Eq.~(7) follows.
Then Eq.~(3) relates $P(s)$ to $\alpha(s)$, where analyticity requires $\alpha(s)$ to be smooth.

In the example of Ishida {\it et al.}, the phase-shift $\delta = \pi$ at
$s=s_0$ of Eq.~(6) and it is essential that this zero is divided out
before constructing the production amplitude, as in Eqs.~(7,8).
In the case of physical $\pi\pi$ scattering,
inelasticity has set in before any phase-shift reaches $\pi$, consequently this example has no relevance
beyond this model of Ishida {\it et al.}.  However, for physical
$\pi\pi$ scattering the determinant of the $T$--matrix does vanish close to
$K{\overline K}$ threshold and defining a reduced ${\hat K}$--matrix eliminates this zero.
This is the multi-channel generalization of the above example discussed 
in Ref.~7.
\vskip 2.5cm

\centerline{\bf References}
\vskip 5mm

\item{   1.  } M.Y. Ishida, S. Ishida and T. Ishida, {\it Relation between scattering
and production amplitudes --- case of intermediate $\sigma$--particle
 in $\pi\pi$ system},
contribution to Hadron'97, Brookhaven,  August 1997.

\item{   2.  } T. Ishida, K. Takamatsu, T. Tsuru, M.Y. Ishida and S. Ishida,
{\it $\sigma$--particle in production processes}, 
contribution to Hadron'97, Brookhaven,  August 1997.

\item{   3.  } S. Ishida, M.Y. Ishida, H. Takahashi, T. Ishida,
K. Takamatsu and T. Tsuru, Prog. Theor. Phys. {\bf 95} (1996), 745.
 
\item{   4.  } K.M. Watson, Phys. Rev. {\bf 88} (1952), 1163.

\item{   5.  } I.J.R. Aitchison, Nucl. Phys. {\bf A189} (1972), 417.

\item{   6.  } K.L. Au, D. Morgan and M.R. Pennington, Phys. Rev. {\bf D35}
(1987), 1633.

\item{   7.  } D. Morgan and M.R. Pennington, Z. Phys. {\bf C48} (1990), 623.

\bye